\documentclass[10pt,preprint2]{aastex}

\shorttitle{Optical counterpart to XTE J1739$-$302}
\shortauthors{Negueruela et al.}

\begin{document}


\title{The optical counterpart to the peculiar X-ray transient  XTE
  J1739$-$302 \altaffilmark{1}}
\altaffiltext{1}{Based on observations collected at the European
  Southern Observatory, La Silla, Chile (ESO 73.D-0081)}


\author{Ignacio Negueruela}
\affil{Dpto. de F\'{\i}sica, Ingenier\'{\i}a de Sistemas y Teor\'{\i}a de 
la Se\~{n}al, Universidad de Alicante, Apdo. 99, E03080 Alicante, Spain
\email{ignacio@dfists.ua.es}}
\author{David M. Smith}
\affil{Physics Department and Santa Cruz Institute for Particle
  Physics, University of California, Santa Cruz, 1156 High St., Santa
  Cruz, CA 95064}
\author{Thomas E. Harrison\altaffilmark{2}}
\affil{Astronomy Department, New Mexico State University,Box
  30001/Dept. 4500, Las Cruces, NM 88003}
\altaffiltext{2}{Visiting
Astronomer at the Infrared Telescope Facility, which is operated by the
University of Hawaii under contract from NASA}
\and

\author{Jos\'e Miguel Torrej\'on}
\affil{Dpto. de F\'{\i}sica, Ingenier\'{\i}a de Sistemas y Teor\'{\i}a de 
la Se\~{n}al, Universidad de Alicante, Apdo. 99, E03080 Alicante, Spain}

\begin{abstract}
The weak X-ray transient XTE J1739$-$302, characterized by extremely
short outbursts, has recently been identified with a reddened
star. Here we present spectroscopy and photometry of the counterpart,
identifying it as a O8\,Iab(f) supergiant at a distance of $\sim2.3\:$kpc.  
XTE J1739$-$302 becomes thus characterized as the prototype of the new
class of Supergiant Fast X-ray Transients. The optical and infrared
spectra of the counterpart to  XTE J1739$-$302 do not reveal any
obvious characteristics setting it apart from other X-ray binaries with
supergiant companions, which display a very different type of X-ray
lightcurve. 
\end{abstract}


%% subject header.  Objects should be in the appropriate "individual"
%% headers (e.g. quasars: individual, stars: individual, etc.) with the
%% additional provision that the total number of headers, including each
%% individual object, not exceed six.  The \objectname{} macro, and its
%% alias \object{}, is used to mark each object.  The macro takes the object
%% name as its primary argument.  This name will appear in the paper
%% and serve as the link's anchor in the electronic edition if the name
%% is recognized by the data centers.  The macro also takes an optional
%% argument in parentheses in cases where the data center identification
%% differs from what is to be printed in the paper.



\section{Introduction}

\defcitealias{smi05}{Paper~I}

The X-ray transient XTE J1739$-$302 = \object{AX J1739.1$-$3020} =
\object{IGR J17391$-$3021} is highly unusual because of its extremely
short outbursts. The only outburst that was monitored over its full
extent was found to last less than one day \citep{lut05a}. In a
companion paper (\citealt{smi05}; henceforth \citetalias{smi05}), we
provide full characterization of the X-ray behavior of XTE
J1739$-$302 and evidence
suggesting that similarly short durations characterize all its
outbursts.

A pointed {\it Chandra} observation led to the unambiguous
identification of the optical counterpart to XTE J1739$-$302. This
star is cataloged as
USNO-B1.0 0596--0585865 = 2MASS 17391155-3020380, and will be referred
to in the following as Star A. Here we report on extensive optical and
infrared observations of this source, a reddened late O-type
supergiant. 

As discussed in \citetalias{smi05}, the interest of XTE J1739$-$302
stems from the fact that there seems to be a large number of similar
objects, characterized by supergiant donors, low quiescence X-ray
luminosities and short ($< 1\:$d) outbursts, which we term
Supergiant Fast X-ray Transients (SFXTs).

\section{Observations}

\subsection{Photometry}

Photometry of star A was obtained on the night of June 27th, 2004,
with the SUperb Seeing Imager (SUSI-2) attached to the 3.5-m  New
Technology Telescope (NTT) at La Silla, Chile. SUSI-2 was equipped
with a mosaic of two $2144\times4096$ EEV CCDs in $2\times2$ binning
mode. The pixel scale is $0\farcs16$/pixel.

Standard Bessell $UBVRI$ filters were used to observe the field of XTE
J1739$-$302 and several Landolt standard fields \citep{landolt}, at a
range of airmasses.  
Bias subtraction and flat fielding were carried
out on all frames using {\em Starlink} {\sc ccdpack} software
\citep{draper}. Then,
using the {\em Starlink} {\sc gaia} software \citep{draper2}, aperture
photometry was 
performed on all frames with background subtraction from annular sky
regions around each star. 

Instrumental minus catalog magnitudes were calculated for each of the
Landolt standard stars and linear fits were performed against airmass. The
resulting extinction coefficient, zero point and color term
corrections were applied to the target star, resulting in the
magnitudes shown in Table~\ref{tab:phot}.  

\subsection{Optical Spectroscopy}

Intermediate-resolution spectroscopy of the source in the red region
was obtained with the ESO Multi-Mode
Instrument (EMMI) on the NTT during a run in June 2003. The red arm
was used in the longslit REMD mode, equipped with a mosaic of two
$2048\times4096$ MIT CCD detectors.

On the night of June 6th, we used grating \#6 to observe the
$\lambda\lambda6440-7140${\AA} range and grating \#7 to observe the
$\lambda\lambda7600-9000${\AA} range. On June 7th,
we only observed XTE J1739$-$302 with 
grating \#7, now centered so as to cover $\lambda\lambda6310-7835${\AA}.

Image pre-processing was carried out with {\em MIDAS} software, while
data reduction was achieved with the
{\em Starlink} packages {\sc ccdpack} and  \citep{draper} and {\sc figaro}
\citep{shortridge}. Analysis was carried out using {\sc figaro} and {\sc dipso}
\citep{howarth}.

The blue part of the spectrum was observed on the night of May 21st,
2004, with the FOcal Reducer/low dispersion Spectrograph (FORS1)
mounted on the 8.2-m VLT/UT1 {\it Antu} telescope. FORS1 was equipped
with a thinned $2048\times2048$ Tektronix CCD and the G600B grism,
which covers the  $\lambda\lambda3600-6000${\AA} range with a nominal
dispersion of 1.2\AA/pixel. The final spectrum, calibrated by the FORS
pipeline, was extracted and analyzed with {\sc figaro} and {\sc
  dipso}. 

\subsection{Infrared Spectroscopy}

\begin{figure}
\epsscale{1.0}
\plotone{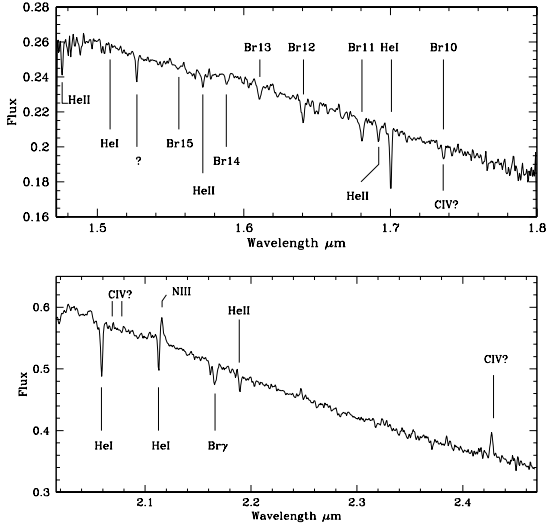}
\caption{Near-IR spectra of XTE~J1739$-$302. The $H$-band spectrum
  (top) shows a prominent \ion{He}{2}~1.693{$\mu$}m, which is only seen
  in O-type supergiants. The $K$-band spectrum shows strong
  \ion{N}{3}~2.116{$\mu$}m emission, characteristic of Of stars
  \citep{han96}. }
\label{fig:irspec}
\end{figure}

Infrared spectroscopy of the counterpart to XTE~J1739$-$302 was
obtained using SPEX on the 
Infrared Telescope Facility (IRTF) on Mauna Kea on 2003 May 17th. The
$H$- and $K$-band spectra presented in Fig.~\ref{fig:irspec}, were
constructed from medians  
of six 10-s exposures. The observing and data reduction procedure are 
fully described in \citet{har04}. SPEX was used in single-order mode 
with a 0.3" slit, giving a dispersion of 5.51~\AA/pix.  Unfortunately,
the conditions at the IRTF were not photometric, but the seeing was
excellent.

We employed a script where data
at six separate positions along the slit were obtained. To remove the
sky background and dark current from each SPEX exposure, we
subtracted the median of the other five exposures obtained in an
observing sequence. This process resulted in six background-subtracted
exposures from which the spectra were extracted using the normal IRAF
methods. The spectra were wavelength calibrated by the extraction of
an argon arc spectrum at the position (aperture) of each spectrum.  
We observed the G2\,V star HD\,150698 (in an identical fashion) to  
correct for telluric features. The procedure for the use of G-type
dwarfs to correct for the telluric features in near-infrared spectra
has been described by \citet{mai96}. G-stars are useful for telluric
correction because they have very few strong absorption lines in the
near-infrared. But simple division of a the spectrum of a program
object by that of an early G-type dwarf does leave residual features
due to \ion{H}{1} absorption and weak metal lines in the G dwarf
spectrum. \citet{mai96} have developed an 
IRAF routine that modifies a high resolution infrared spectrum of the
Sun for the radial and rotational velocities of the telluric
standard. It then smooths this spectrum to the resolution of the
spectrograph used to observe the telluric standard. We followed their
procedure to correct our data for Star A.

\section{Results}

\subsection{Spectrum description and spectral classification}

\begin{figure}
\resizebox{\hsize}{!}{\includegraphics[bb=160 175 440 675, angle=-90]{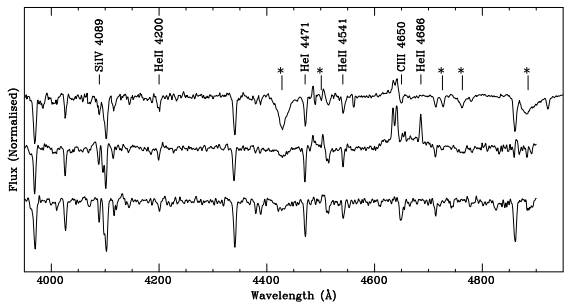}} 
\label{fig:blue} 
\caption{Classification spectrum of the optical counterpart to
  XTE~J1739$-$302 (top). The prominent DIBs are marked with '*'. The
  emission lines are \ion{S}{4}~4486, 4504\AA\ and a \ion{N}{3} blend
  around $\lambda4640$\AA. Also shown for comparison are the O8\,Iaf
  supergiant HD~151804 (middle) and the O8.5\,Iab(f) star HD~112244,
  taken from the digital atlas of \citet{wf90}.} 
\end{figure}

The classification spectrum of the optical counterpart to XTE
J1739$-$302 is displayed in Fig.~\ref{fig:blue}. The presence of
prominent \ion{He}{2} lines identifies it as an O-type star. The
spectral type of O-type stars is determined by the ratio between
\ion{He}{1}~4471\AA\ and \ion{He}{2}~4541\AA, which is unity at
O7 \citep[cf.][for
  classification criteria]{wf90}. The observed ratio indicates an O8
spectral type.  

The presence of
strong \ion{N}{3} emission and neutralization of \ion{He}{2}~4686\AA\
are typical of O-type supergiants. However, the counterparts to
High-Mass X-ray Binaries (HMXBs)
sometimes display emission lines arising from the vicinity of the
compact object. For instance, \object{HD 226868},
the optical counterpart to Cyg X-1, displays transient
\ion{He}{2}~4686\AA\ emission and a variable emission component in
H$\alpha$ which move in antiphase with the photospheric lines
\citep{nin87}. 

\begin{figure}
\resizebox{\hsize}{!}{\includegraphics[bb=130 120 415 710, angle=-90]{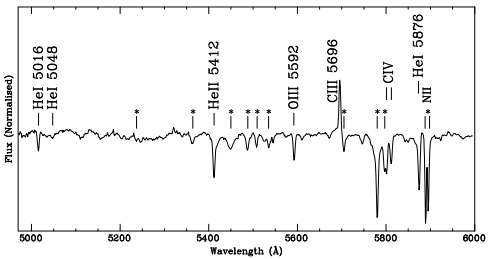}} 
\label{fig:vband} 
\caption{Spectrum in the yellow region of the optical counterpart to
  XTE~J1739$-$302. The strong \ion{C}{3}~5696\AA\ emission is typical
  of late O-type supergiants. The most prominent interstellar DIBs
  seen in the spectrum of 4U~1907+09 by \citet{cox05b} are marked with a
  '*'.} 
\end{figure}

Star A, however, can be safely classified as a supergiant, based on a
large number of characteristics, such as the strong
\ion{S}{4}~4486, 4504\AA\ emission lines or very strong \ion{C}{3}~5696\AA\
emission (see Fig.~\ref{fig:vband}). Moving to absorption features,
the strong \ion{Si}{4}~4089 
\& 4116\AA\ absorption
lines clearly identify the object as a supergiant .

This classification is corroborated by the $H$ and $K$-band spectra,
which can be seen in Fig.~\ref{fig:irspec}. The sharp 
and strong Brackett lines, as well as the sharp and 
prominent \ion{He}{1}~1.700$\mu$m line clearly identify the star as a
supergiant, while the relatively strong \ion{He}{2} lines at
1.693, 2.1885$\mu$m confirm that the star is earlier than
O9. Supergiant Of stars displaying \ion{N}{3}~4640{\AA} in emission
also display \ion{N}{3}~2.116$\mu$m strongly in emission
\citep{han96}. This line is prominent in Star A, though Br$\gamma$ is in
absorption. \citet{han96} show that late Of stars may show  Br$\gamma$
in emission or in absorption, while supergiants earlier than O7 always
show it in emission.

\begin{figure}
\resizebox{\hsize}{!}{\includegraphics[bb=130 120 415 710, angle=-90]{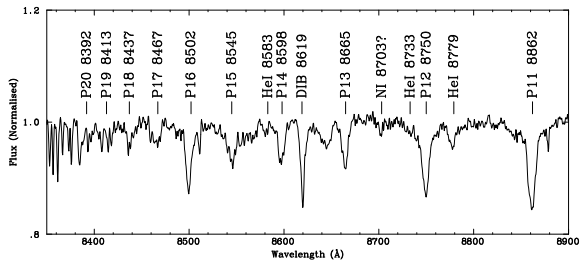}} 
\label{fig:iband} 
\caption{$I$-band spectrum of the optical counterpart to
  XTE~J1739$-$302. } 
\end{figure}

The supergiant classification is also borne out by the $I$-band
spectrum, shown in Fig.~\ref{fig:iband}. Among early-type stars narrow
and deep Paschen and \ion{He}{1} lines can only be seen in luminous
supergiants \citep{car03,mt99}.

\begin{figure}
\resizebox{\hsize}{!}{\includegraphics[bb=85 300 500 625]{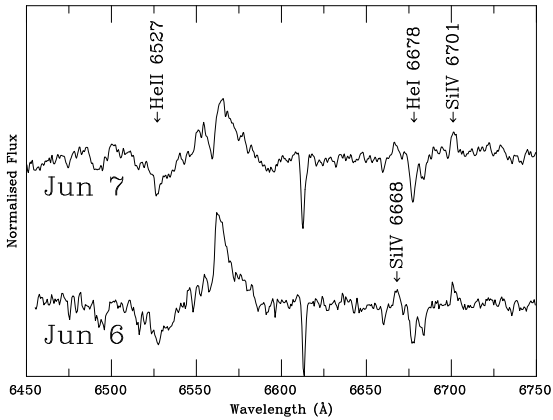}} 
\label{fig:alpha} 
\caption{Intermediate resolution spectra of Star A in the H$\alpha$
  region, taken with the NTT on two successive nights in 2003
  June. Notice the important changes in the shape of H$\alpha$ and the
  intensity of \ion{He}{1}~6678\AA\ (easily comparable with the
  neighbouring \ion{He}{2}~6683\AA). Such changes are completely
  unexpected in a single O-type star, but typical of HMXBs.  } 
\end{figure}

The exact luminosity class of the optical components of HMXBs is
difficult to establish, as the selective emission lines, which are
very sensitive to luminosity, may present emission components not
originating 
from the massive star, likely due to a photoionization wake
trailing the X-ray source \citep{kap94}. Fig.~\ref{fig:alpha} shows
the shape of the H$\alpha$ line in Star A on the nights of
June 6th and 7th, 2003. The obvious variability on such short
timescale would be quite unexpected on an isolated star. It resembles
strongly changes observed in the H$\alpha$ line of \object{HD
  226868}. Very similar changes are also observed in the optical
counterpart to \object{4U~1907+09}, which is likely also an  O8\,I supergiant
\citep{cox05a}.

 In Fig.~\ref{fig:blue}, we compare the spectrum of Star A to those of
 two stars of similar spectral type, taken from the digital atlas of
 \citet{wf90}, the O8\,Iaf
  supergiant HD~151804 and the O8.5\,Iab(f) star
 HD~112244. Star A appears to be more luminous than HD~112244, but
 clearly less luminous than HD~151804. The \ion{He}{2}~4686\AA\ line
 needs to be in emission in order for a star to qualify as Of. In Star
 A, it is only neutralized, which corresponds to an (f) subscript. So
 the object is most likely O8\,Iab(f).

\subsection{Reddening and distance}

For an O8\,I star, \citet{weg94} lists $(B-V)_{0}=-0.30$. Therefore,
the observed $(B-V)=2.97$ implies $E(B-V)=3.27$. The object is
therefore very reddened. For such high reddenings, the photometric
transformations can be subject to errors due to hidden color terms
owing to the very different spectral energy distribution (SED) of the
standard stars and the target. We attempted to correct for this effect
by including among the standards some very red stars from Landolt's
selected area SA~110. 

From our observed colors and the intrinsic colors for an O8\,I star, we find
$E(U-B)/E(B-V)=0.87$. This value is higher than the
standard $0.70$, suggesting that the reddening to XTE
J1739$-$302 is not standard. 
In order to derive a good estimate of the
reddening and distance, we used the $\chi^{2}$ code for parameterized
modeling and characterization of photometry and spectroscopy
implemented by \citet{maiz04}. 

We used as input the SED of a {\sc tlusty} stellar model with
parameters $T_{{\rm eff}}=32500\:$K and $\log g =3.25$, roughly
corresponding to an O8\,I supergiant according to the
observational calibration of \citet{martins05}, our 
optical photometry and the $JHK_{s}$ photometry from 2MASS. The
program attempts to fit an extinction law from \citet{car89} to the
data. In order to allow for hidden color terms, we run the code both
with the photometric errors and adding an extra $0.1\:$mag error to
each optical band. The results were almost indistinguishable. The best
fit is obtained for $E(B-V)=3.37\pm0.06$, $R=2.81\pm0.07$. Adopting
these  values, the code derives a dereddened $m_{V}=5.46$. For an
O8\,I supergiant, \citet{martins05} give $M_{V}=-6.3$, which implies
$DM=11.8$, corresponding to a 
distance of $2.3\:$kpc. The error in this distance is most likely to be
dominated by the uncertainty in the $M_{V}$ calibration. For realistic
  errors of $0.5\:$mag in the $DM$, the distance falls in the range
  $1.8-2.9\:$kpc. Therefore XTE~J1739$-$302 is certainly much closer
  than the distance to the Galactic Bulge.    

The reddening to isolated stars can also be estimated from the
strength of the Diffuse Interstellar Bands (DIBs) in its spectrum. A
good number of prominent DIBs are seen in our spectra of Star
A. However, the relationship between the strength of the interstellar
bands and $E(B-V)$ is not well established for values of the reddening
as high as seen in Star A, mainly because of the scarcity of data
points. 

According to the recent study by \citet{cox05b} most bands present a
non-linear behavior in this very-high-reddening regime. We note in
pass that the interstellar DIBs in the spectrum of Star~A are
comparable in strength to those of the counterpart to 4U~1907+09,
which \citet{cox05b} claim to be stronger than in any other spectrum
observed to date. Along the light
of sight studied by \citet{cox05b}, only two interstellar bands,
those at 5780\AA\ and 5797\AA\ seem to follow the linear
relationship seen at lower reddenings. We cannot use 5797\AA\ because,
in our spectrum,
it appears blended with at least another DIB at 5795\AA\ and the
photospheric \ion{C}{4}~5801 \& 5812\AA\ lines (see
Fig.~\ref{fig:vband}).  This 
leaves us with the 5780\AA\ line as the only trustworthy indicator of
interstellar reddening. Applying the calibration of \citet{her93}, we
find $E(B-V)=3.4\pm 0.2$ (where the error comes mainly from the
uncertainty in the equivalent width of the line). This value agrees
surprisingly well with the spectrophotometric determination.

\section{Discussion}

Until recently, it was widely believed that High-Mass X-ray Binaries
(HMXBs) with supergiant donors were always persistent X-ray sources,
while Be/X-ray binaries were mostly transients, though some of them
were persistent low-luminosity X-ray sources
\citep[cf.][]{rr99}. Supergiant HMXBs are believed to be powered by
direct accretion from the strong radiative wind of the supergiant.
They display typical X-ray luminosities in the $L_{\rm X}\sim 
10^{35}-10^{36}\:{\rm erg}\,{\rm s}^{-1}$ range, which are relatively constant
over the long run, 
though presenting strong short-term stochastic variability \citep{km05}.

The behavior of XTE~J1739$-$302 differs from this pattern in two
important respects. On the one hand, its quiescent X-ray luminosity is rather
lower than this  ($\la 10^{34}\:{\rm erg}\,{\rm s}^{-1}$) and the
lightcurve shows abrupt ``outbursts'' lasting only several hours. 
As discussed in \citetalias{smi05}, several other X-ray sources display
similar short outbursts and have OB supergiants as mass donors. We
have termed this class of objects SFXTs.

A third interesting property of XTE~J1739$-$302 is the fact that the
absorption column density derived from the fits to X-ray spectra is
not only larger than the expected interstellar absorption, but also
highly variable. The optical extinction to the source
$E(B-V)=3.4$ transforms to a $N_{\rm H} = 2\times10^{22}\:{\rm
  cm}^{-2}$. In contrast, the hydrogen column
densities measured at different times range from $N_{\rm H} = 3.2$ to
38 $\times10^{22}\:  
{\rm cm}^{-2}$ \citepalias{smi05}, clearly suggesting that most of the
absorption seen in 
the X-ray spectra is intrinsic to the source.

Similarly variable absorption has been observed in a number of sources
recently detected by the {\it INTEGRAL} satellite. These objects are
generally very absorbed, resulting in very hard X-ray spectra
\citep{wal04}, which explain the lack of detections by previous
missions.  These objects have been dubbed ``obscured {\it INTEGRAL}
sources'' \citep[see, e.g.,][]{kuul05}.

Many of these new {\it INTEGRAL} sources display X-ray pulsations and
are hence HMXBs, while several others are also likely to be HMXBs
\citep[cf.][]{lut05b}, among them some with rather unusual properties
\citep[e.g.,][]{fc04}. Two of the new {\it INTEGRAL} sources,
IGR~J17544$-$2619 and IGR~J16465$-$4507 have already been shown to come
within the SFXT category \citepalias{smi05}. However, there does not
appear to be a direct
correlation between high obscuration and fast outbursts. As a matter
of fact,
IGR~J17544$-$2619, which also has as counterpart a late O-type
supergiant \citep{pell05}, seems to show lower and less variable
absorption than XTE~J1739$-$302 \citep[cf.][though this could well be
  due to the limited number of observations]{gr04}.

There is nothing in the optical/IR spectrum of Star A that can provide
an explanation to its X-ray behavior. In the optical, XTE~J1739$-$302
is very similar to the wind-fed 
supergiant system 4U~1907+09 \citep{cox05a}, which has a donor of
approximately the same spectral type and is extincted by a similar
amount. 4U~1907+09 is believed to be at $d\ga5\:$kpc and it is so more
than one magnitude fainter than Star A. The X-ray properties of
4U~1907+09 and XTE~J1739$-$302 are, however, completely
different. 4U~1907+09 is a persistent X-ray source showing increased
emission every $8.4\:$d, in correspondence with the periastron passage
of a neutron star, which is in a relatively eccentric orbit
($e=0.28$). Its average X-ray luminosity $L_{\rm X}\sim 
10^{36}\:{\rm erg}\,{\rm s}^{-1}$ \citep{cox05a} is typical of wind-fed HMXBs.

The exact shape of the lightcurve in a wind-fed HMXB depends on many
parameters, perhaps most fundamentally on the wind structure
(determined by the evolutionary status of the supergiant) and the
orbital eccentricity. 4U~1907+09 and GX~301$-$2 show strong peaks near
periastron, while other systems have smoother lightcurves. But none of the
wind-fed HMXB studied in detail so far has properties similar to those
of XTE~J1739$-$302.

In principle, one
could perhaps attribute the very low quiescence luminosity of
XTE~J1739$-$302 to a very wide orbit, but the strong variations in the
shape of H$\alpha$ between two consecutive nights would support a
relatively close interaction, as (at least one component of) H$\alpha$
should be tracing the vicinity of the compact
object. In any case, it seems that the exact meaning of a ``quiescent
state'' is difficult to define for XTE~J1739$-$302 \citepalias{smi05} and
IGR~J17544$-$2619 \citep{gr04,zand05}. Furthermore, the
detection of possible fast outbursts 
from Vela X-1 and Cyg X-1 rules out a correlation between low X-ray
luminosity and fast outbursts or between a particular type of compact 
object and fast outbursts, strongly suggesting that the physical
mechanism behind the fast outbursts is related to the accretion of
material from the wind of a supergiant.

%As discussed in \citetalias{smi05}, several other transients display
%behaviours similar to XTE~J1739$-$302. In particular, AX~1845.0$-$0433
%and IGR~J17544$-$2619 both have as counterparts late O-type
%supergiants \citep{coe96,pell05}. The latter object, which presents a
%moderately absorbed  X-ray spectrum as is not very obscured in the
%optical,  level. We may suspect them

Whatever the cause for this behavior, XTE~J1739$-$302 is the first
member of a new class of HMXB to be fully characterized. The fact that
it has only been detected because it happens to be very close to a
frequently observed source opens the possibility that a substantial
population of similar low X-ray luminosity objects exists.

\acknowledgments

 IN  is a researcher of the
program {\em Ram\'on y Cajal}, funded by the Spanish Ministerio de
Ciencia y Tecnolog\'{\i}a and the University of Alicante, with partial
support from the Generalitat Valenciana and the European Regional
Development Fund (ERDF/FEDER). 
This research is partially supported by the Spanish MCyT under grants
AYA2002-00814 and ESP-2002-04124-C03-03. 

Part of the data used here have been obtained through the service
program of the European Southern Observatory. We thank the staff
astronomers for their dedication. We also thank Dr.~J.~Simon Clark and
Dr.~Amparo Marco for their
help during the 2003 NTT run and Dr.~Jes\'us Ma\'{\i}z-Apell\'aniz for
help with the use of the {\sc chorizos} code. 

This publication makes use of data products from the Two Micron All
Sky Survey, which is a joint project of the University of
Massachusetts and the Infrared Processing and Analysis
Center/California Institute of Technology, funded by the National
Aeronautics and Space Administration and the National Science
Foundation. 
 
Facilities: \facility{VLT(FORS1)}, \facility{NTT(SUSI2)},
\facility{NTT(EMMI)},\facility{CXO(ASIS)}.

\clearpage

\clearpage

\clearpage

\begin{table}
\begin{centering}
%\tablewidth{0pt}
\caption{Photometric values for the optical counterpart to
 XTE J1739$-$302 derived from
 our observations. \label{tab:phot}}
\begin{tabular}{ccccc}
\tableline
$U$& $B$&$V$ &$R$ &$I$\\
\tableline
&&&&\\
$19.54\pm0.04$&$17.86\pm0.02$&$14.89\pm0.02$&$12.96\pm0.02$&$11.35\pm0.03$\\
\tableline
\end{tabular}
\end{centering}
\end{table}

\begin{table}
\begin{centering}
%\tablewidth{0pt}
\caption{IR photometric values for the optical counterpart to
 XTE J1739$-$302 from
 2MASS.\label{tab:irphot}}
\begin{tabular}{ccc}
\tableline
$J$& $H$& $K_s$\\
\tableline
&&\\
$8.600\pm0.021$&$7.823\pm0.021$&$7.428\pm0.017$\\
\tableline
\end{tabular}
\end{centering}
\end{table}

\end{document}